# Precise design of VO$_2$ thin films for smart windows by employing thickness dependent refractive index

*Mohammad Hossein Mahdieh, Mehrad Sohrabi

Department of Physics, Iran University of Science and Technology

Narmak, Tehran, Iran

**Abstract.** Vanadium dioxide (VO$_2$) is an adjustable refractive index material and has capability of behaving as semiconductor or conductor depending on its temperature. Such condition makes it as a material which can be employed in fabricating thermochromic smart windows. The transmission characteristics of these type of windows strongly depend on the thickness of the film. Therefore, some calculations are required to optimize the VO$_2$ thickness. Unfortunately, refractive index of VO$_2$ thin film is thickness dependent, therefore, in calculating the transmission of light spectrum from VO$_2$ thin films, a unique refractive index cannot be utilized. In the present paper using three theoretical models (Lorentz-Drude oscillator, Lorentz oscillator, and Tauc-Lorentz) and employing experimental results from previous reports, we provide a collection of thickness dependent refractive index of VO$_2$ films. More precise transmission can be achieved by using this set of refractive index data in the calculations which agree with those of experiments. These results also fairly make us capable to determine optimized film thickness for any desired transmission performance. This method is useful for design of VO$_2$ based thermochromic smart windows.

## 1 Introduction

Energy consumption in buildings, transportations, and other industries has led to growing the idea of using renewable energy and also energy saving.[1-3] Particular attention has been paid to energy saving in buildings since it consumes about 30% to 40% of the main energy in the world.[1, 4] Therefore, energy saving has become a major issue in building design and construction.

Windows in buildings have the main important role in energy saving. An appropriate way to make an effective use of solar energy in buildings is to modulate sunlight transmission through glass windows by optical technique that is called "smart windows".[5, 6] Smart window, as a common method for energy saving in buildings has been proposed for more than four decades. Within these periods, the design of smart windows has been improved by utilizing chromogenic materials.[7, 8] In moderate climates, it is desirable that the optical properties of such window coatings (chromogenic coatings) vary so that the material controls the transmission (or reflection) of light spectrum depending on the conditions of the environment temperature.

Chromogenic materials have been designed in different types to control the transmitted light in smart windows. For example, electrochromic chromogenic materials (EC) can control the transmitted light by responding to an external voltage.[9-11] Thermochromic chromogenic materials (TC) control the entrance light by variation of temperature.[12, 13] In addition, there are other

---

* Address all Correspondence to: Mohammad Hossein Mahdieh, E-mail: mahdm@iust.ac.ir

chromogenic alternatives such as photochromic (PC)[14, 15] which become opaque under irradiation. Gasochromic (GC) materials control the light using gaseous and the transmission in GC depends on the gas.[16, 17]

Among chromogenic materials, thermochromic (TC) has been most interested since it can be utilized without extra input energy and can regulate the transmission of solar spectrum via adaptive response to environment temperature.[18] The TC is potentially low-cost relative to the other technologies due to its fabrication simplicity.

Vanadium dioxide ($VO_2$) has been known as the best TC material for energy saving in smart windows.[19, 20] At a critical temperature ($\sim 68\ ^0C$), a reversible semiconductor to metal transition takes place in bulk $VO_2$.[21] Such phase transition properties makes $VO_2$ a suitable material for fabricating varieties of electrical and optical devices, including uncooled infrared bolometers, TC coatings in smart windows, fiber optical switching devices, and optical memory elements.[22]

As explained, the transmission in smart windows coated with $VO_2$ can be controlled if the ambient temperature varies. The light transmission through such windows depends on the ambient and a critical temperature. In fact, in a window which is coated with chromogenic $VO_2$ film, when the room temperature is lower than the critical temperature, both visible and near-infrared light have high transmittance. In such case the temperature in the room gradually increases with the high entrance of infrared light. However, when the room temperature is higher than the critical temperature, visible light transmittance is maintained high but near-infrared transmittance is significantly decreased. In such condition, the infrared light is blocked and heat generated by the infrared spectrum cannot enter the room. The reason for such functioning is that during the phase transition, the crystal structure of vanadium dioxide changes from monoclinic ($P2_1/c$, M1)[23] to rutile state ($P4_2/mnm$ R).[24]

Phase transition in $VO_2$ thermochromic chromogenic makes it a good candidate for smart windows, however, there are some limitations in this material which need to be considered for having better performance. The main shortcomings of the $VO_2$ thermochromic chromogenic are its low transmission rate for visible light ($T_{lum}$), unpleasant coating colors, high critical temperature ($T_t$), and low solar modulation ability $\Delta T_{sol}$. Solar modulation ability $\Delta T_{sol}$ (and will be defined the next section) is the difference in solar-energy transmittance before and after the phase transition.[25-28] Fortunately, the above factors can be controlled as they are mainly dependent on the microstructure, thickness, optical constant of the coating, and fabricating technique.[29]

Konovalova et al. have shown that the optical constants of $VO_2$ films largely depend on their fabrication technology.[30, 31] Different techniques have been utilized to fabricate the thermochromics material to obtain appropriate condition for its performance. For example, $VO_2$ thin films were deposited on amorphous silica substrates with a specific preferential crystal orientation by RF reactive sputtering process.[32] By this technique, different samples with thicknesses ranging from 30 nm to 200 nm were fabricated. It was shown that maximum optical transmittance was observed for 120 nm of the $VO_2$ film. For a single wavelength good agreement between experimental and calculating data was observed just for thicknesses greater than 80 nm. The discrepancy between the experimental and calculation results are mainly due to inappropriate calculating model, or (and) utilization of imprecise refractive index in the calculations. For example, simulation by FDTD Solution software for $VO_2$ films (with thicknesses ranging from 30 nm to 120 nm) show that only for thinner films, the calculation results are in good agree with those of experiment.[33]

Measurement of $T_{lum}$ and $T_{sol}$ were performed for VO$_2$ films (ranging from 25 nm to 110 nm) and the results were compared with those of calculations.[34] This study showed that only for 50 nm thick film the experimental and calculation results agree each other but there was significant difference for the samples with thicknesses thicker than 50 nm. The main reason for such discrepancy was also using inappropriate data of refractive index of samples with thicker films. In another research various thicknesses (30-540 nm) of polycrystalline VO$_2$ thin films were deposited on Borosilicate Crown glass (BK7) and the refractive index of the films were measured. This study confirmed the disagreement between calculations and experimental results and showed substantial discrepancy between the experimental and calculation results. Such significant differences were also due to using improperly model for refractive index.[34] The spectral transmission characteristics of windows was attempted to be enhanced by adding a quarter wave optical thickness TiO$_2$ film on VO$_2$. By this technique, the integrated luminous reflectance ($R_{lum}$) of VO$_2$ was reduced dramatically, however, the simulation data did not support the experiment results.[35] The results of an investigation indicated that special design which was based on optimized optical thicknesses can perfectly improve the visible transmittance and the solar modulation efficiency.[36] However, the results of this report were just for a specific thickness and not comprehensive In fact, in all the above mentioned reports, an inclusive data set for the refractive index of VO$_2$ film (thickness dependent refractive index) was not presented.

In the present research, the transmission of VO$_2$ film was evaluated numerically by utilizing thickness dependent refractive index comprehensively. Using the most appropriate thickness dependent refractive index, not only the simulation results can support previous experimental results but also such data was used for finding optimized condition for the best transmission of VO$_2$ with any thickness.

## 2 Calculation methods

### 2.1 Complex dielectric functions

As explained in the previous section, VO$_2$ is utilized for fabricating smart windows due to its adjustable refractive index in metal and semiconductor phases. Furthermore, the refractive index of vanadium dioxide films vary with thickness and its bulk data cannot be used for VO$_2$ thin film structures. Practically, the complex refractive index of VO$_2$ films mostly was measured by ellipsometric technique, in both the semiconductor and the metallic states. In calculating optical transmission, it is important to choose correct values (that is thickness dependent) for this parameter or else the calculating results do not agree with those of the experiment. Therefore, in order to find realistic results, using suitable model and data for calculating refractive index become crucial issues.

There are several dispersion models which express the dielectric function of VO$_2$. In general, the complex refractive index of VO$_2$ film can be described mainly by three models, i.e. Lorentz–Drude oscillator, Lorentz oscillator, and Tauc-Lorentz. Equations 1 to 3 describe these three models.

   *i)*      Lorentz-Drude oscillator

$$\varepsilon(\omega) = \varepsilon_\infty - \frac{\omega_P^2}{\omega^2 + i\omega_c\omega} + \sum_{j=1}^{n} \frac{f_j}{1 - \omega^2/\omega_j^2 - i\gamma_j\omega/\omega_j} \quad (1)$$

Where, $\omega_p$ is plasma frequency, $\omega_c$ is collision frequency of the free electrons, $\omega_j$ is phonon resonance frequency, $f_j$ is strength of oscillator, and $\gamma_j$ is the line width.

*ii)* Lorentz oscillator

$$\varepsilon(E) = \varepsilon_\infty + \sum_{j=1}^{n} \frac{A_j}{(E_j^2 - E^2) - iB_jE} \quad (2)$$

In equations 1, and 2, $\varepsilon(E)$ is the complex dielectric permittivity as a function of photon energy, $E$. The permittivity is a function of the high frequency permittivity $\varepsilon_\infty$ plus a sum of $n$ oscillators where $A_j$ is the oscillator amplitude, $E_j$ the oscillator energy, and $B_j$ the oscillator damping.

*iii)* Tauc-Lorentz

$$\varepsilon(E) = \varepsilon_1(E) + i\,\varepsilon_2(E)$$

$$\varepsilon_1(E) = \varepsilon_\infty + \frac{2}{\pi} p \int_{E_g}^{\infty} \frac{\xi \varepsilon_2(\xi)\,d\xi}{\xi^2 - E^2}$$

$$\varepsilon_2(E) = \begin{cases} \frac{1}{E} \frac{AE_0 C(E - E_g)^2}{(E^2 - E_0^2)^2 + C^2 E^2}, & E > E_g \\ 0, & E \leq E_g \end{cases} \quad (3)$$

Where, $\varepsilon_\infty$ represents the value of real part of the dielectric function $\varepsilon$ at infinite energy (an additional fitting parameter in the oscillator model), $p$ stands for the Cauchy principal part of the integral. Also, $E_g$ is the bandgap of the material, $A$ is the oscillator amplitude, $E_0$ is the energy of the Lorentz peak, and $C$ is the broadening parameter.

The research results showed that different thickness can influence the optical characteristics of VO$_2$ film.[37-44] In each of these references, the refractive index of a single VO$_2$ thin film was determined by calculation and experimental measurement. The measurement was usually performed by ellipsometry technique and then using one of the above models, the calculated refractive index was fit with the experiment by tuning the parameters in the equations. However, not necessarily using the calculated refractive index is precise enough to be used for calculating the transmission of light through a film with arbitrary thickness. The transmission of VO$_2$ films with variety of thicknesses are also available in references.[34, 35, 45-51] However, the calculated transmission in references[34, 35, 45-51] do not support completely those of the experiment.

In the present work by using the above models, we generated the wavelength dependent and thickness dependent complex refractive index of VO$_2$. The parameters in these equations were taken from the data in references[37-44] that are introduced in Table 1. In Table 1, for simplicity the data set from references are recognized by specific thickness and identified by M(x) where "x" stands for the specific thickness that was used in each particular reference. This approach provides thickness dependent data for refractive index which now can be used to achieve the best transmission for any preferred film thickness. Such data have not been available as a package so far. In this paper we have provided such availability by performing the following steps. We initially used these refractive index data to calculate the transmission of electromagnetic spectrum through $VO_2$ films with various thicknesses. The transmission of these particular $VO_2$ films were already measured experimentally and addressed in references.[34, 35, 45-51] Then the calculated transmissions were compared with those of experiments. Finally, if the calculated results were in good agreement with those of experiment, those specific data set was labeled as the best set for that specific film thickness. By this approach, appropriate models for calculating thickness dependent refractive index were recognized. Generating such refractive index data are very useful for smart windows design. The methodology can be used as a guide to minimize the discrepancy between the calculation results and those of experiments. This approach (and the results which have been achieved) was not considered extensively in a single paper in previous researches.

**Table 1** The information of data set M(x) in which "x" is the thickness. The required parameters (such as $\omega_p$ plasma frequency, $\omega_c$ collision frequency of the free electrons, $\omega_j$ phonon resonance frequency, $f_j$ strength of oscillator, and $\gamma_j$ damping coefficient for each oscillator) for calculating thickness dependent refractive index (equations 1 to 3) were extracted from the indicated references [37-44].

| M(x) "x" is the VO₂ film thickness | Reference | Dispersion model |
|---|---|---|
| M(35) | 37 | Lorentz Oscillator |
| M(52) | 38 | Lorentz-Drude oscillator |
| M(60) | 39 | Lorentz-Drude oscillator |
| M(69) | 40 | Lorentz-Drude oscillator |
| M(79) | 41 | Lorentz oscillator |
| M(80) | 42 | Lorentz oscillator |
| M(100) | 43 | Lorentz-Drude oscillator |
| M(109) | 40 | Lorentz-Drude oscillator |
| M(113) | 40 | Lorentz-Drude oscillator |
| M(134) | 40 | Lorentz-Drude oscillator |
| M(180) | 44 | Tauc-Lorentz oscillator |
| M(B) | 43 | Lorentz-Drude oscillator |

## 2.2 Transmission of the film

The transmission through a thin film (deposited on a substrate) can be easily calculated by Fresnel equations if the refractive index of film, and substrate are known.[52] A straightforward set of formulae for our case (a VO₂ thin film deposited on a glass substrate) is given in reference [53] and detailed relations are described by relations 4. In these relations, $T$ is the transmission, $\lambda$ is the light wavelength, $d$ is the film thickness, $n_1$, and $k_1$ are the real and imaginary parts of refractive indexes of the film and $n_2$, and $k_2$ are the real and imaginary parts of refractive indexes of the substrate respectively. Furthermore, the device (the substrate and the deposited VO₂ film) is assumed to be held in air environment (with index of refraction of $n_0 \sim 1$).

$$T = \frac{n_2}{n_0} \frac{\{(1+g_1)^2 + h_1^2\}\{(1+g_2)^2 + h_2^2\}}{\exp(2\alpha_1) + (g_1^2 + h_1^2)(g_2^2 + h_2^2)\exp(-2\alpha 1) + A_H \cos 2\gamma_1 + B_H \sin 2\gamma_1} \quad (4)$$

$$A_H = 2(g_1 g_2 - h_1 h_2) \qquad B_H = 2(g_1 h_2 + g_2 h_1)$$

$$g_1 = \frac{n_0^2 - n_1^2 - k_1^2}{(n_0 + n_1)^2 + k_1^2} \qquad g_2 = \frac{n_1^2 - n_2^2 + k_1^2 - k_2^2}{(n_1 + n_2)^2 + (k_1 + k_2)^2}$$

$$h_1 = \frac{2 n_0 k_1}{(n_0 + n_1)^2 + k_1^2} \qquad h_2 = \frac{2(n_1 k_2 - n_2 k_1)}{(n_1 + n_2)^2 + (k_1 + k_2)^2}$$

$$\alpha_1 = \frac{2\pi k_1 d}{\lambda} \qquad \gamma_1 = \frac{2\pi n_1 d}{\lambda}$$

Further two parameters which are required to be introduced for this investigation are $T_{lum}$ and $T_{sol}$. The integrated values of luminous transmittance ($T_{lum}$, 380 to 780 nm), and solar transmittance ($T_{sol}$, 300 to 2500 nm) which are used to evaluate the performance of smart windows are defined by equations 5, and 6 respectively.[35, 54]

$$T_{lum} = \frac{\int_{380}^{780} \Phi_{lum}(\lambda) T(\lambda) \, d\lambda}{\int_{380}^{780} \Phi_{lum}(\lambda) \, d\lambda} \quad (5)$$

$$T_{sol} = \frac{\int_{300}^{2500} \Phi_{sol}(\lambda) T(\lambda) \, d\lambda}{\int_{300}^{2500} \Phi_{sol}(\lambda) \, d\lambda} \quad (6)$$

where $\phi_{lum}$ is the spectral sensitivity of the light-adapted eye and $\phi_{sol}$ is the solar irradiance spectrum for an air mass of 1.5, $T(\lambda)$ is the transmittance at wavelength $\lambda$.[55] The solar irradiance spectrum for an air mass is defined by "The American Society for Testing and Materials" (ASTM) which represents terrestrial solar spectral irradiance on a surface of specified orientation under one and only one set of specified atmospheric conditions [56]. Such distributions of power (watts per square meter per nanometer of bandwidth) as a function of wavelength provides common criteria for assessing spectrally selective materials with respect to performance measured under varying natural and artificial sources of light with various spectral distributions.

## 3 Results and discussions

Using equations 1 to 3, the real and imaginary parts of the complex dielectric permittivity (and therefore refractive index) of VO$_2$ versus wavelength (300 nm to 2500 nm) in semiconductor and conductor phases were calculated. As explained in previous section, the values of physical parameters (i.e. $\omega_p$ plasma frequency, $\omega_c$ collision frequency of free electron, $\omega_j$ phonon resonance frequency, $f_j$ strength of oscillator, and $\gamma_j$ damping coefficient for each oscillator) in equations 1 to 3 were extracted from references in Table 1. In each of these references, one of these three models was used for a specific film thickness. Therefore, we identified the calculated refractive index by label M(x) where "x" is the film thickness and described in Table 1. The results of these calculations are presented in Fig. 1.

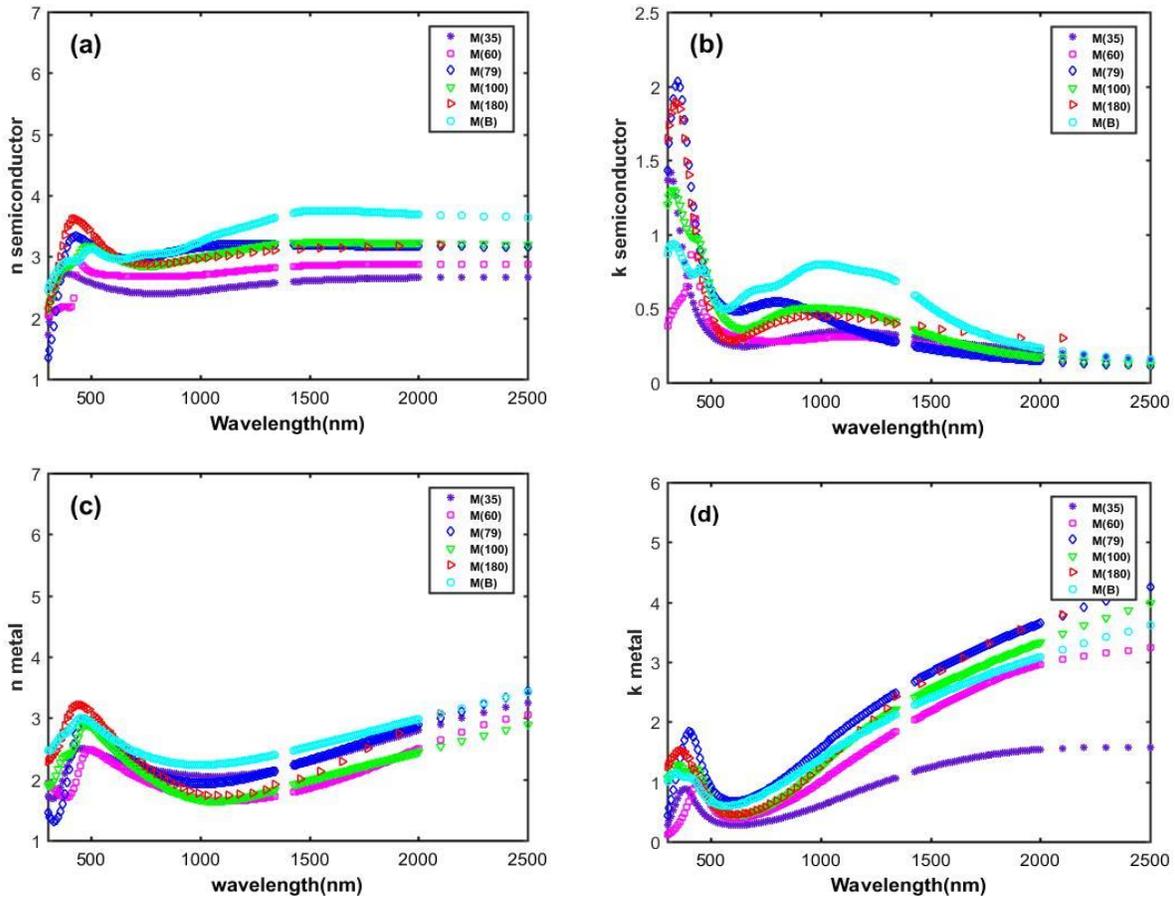

**Fig. 1** Refractive index of VO$_2$ film using data from references **(a)** real part of refractive index of semiconductor phase **(b)** imaginary part of refractive index at semiconductor phase, **(c)** real part of refractive index of metallic phase, and **(d)** imaginary part of refractive index at metallic phase

The data in this figure show that the refractive index closely depends on the thickness. Furthermore, these results show that the calculated refractive index strongly depends on two main issues: ***i*)** the model (one of the equations 1 to 3), and ***ii*)** the parameters in there which are reported in different references in Table 1. Such differences in the calculation results (data in Fig. 1) are mainly associated with the dissimilar values (or incorrect values) for the optical parameters which were addressed in references in Table 1.

As explained the parameters (which we have used in equations 1 to 3) were extracted from data which are belong to just some specific film thickness and reported in references in Table 1. Therefore, there is no guarantee that these extracted values are accurate enough so that if we use them in the calculation of refractive index, necessarily give correct results for any other arbitrary film thickness. Consequently, if one uses these data may not get the right transmission which fit with those of experiments. This is why different values for refractive index may be achieved for even similar condition (Fig.1). Therefore, it is required to find correct data for thickness dependent refractive index by comparing all available experimental records for $VO_2$ transmission with our calculated transmission which are based on thickness dependent refractive index.

Regardless of the differences of the results in the Fig. 1(a) and 1(b), it can be seen that the trend of both imaginary and real parts of the refractive index of semiconductor phase increase by increasing the thickness. The variations of these two parameters for conducting phase are similar for almost all thicknesses. There are significant lack of data which are related to the complex refractive index of $VO_2$ films for thicknesses larger than approximately 150 nm. Hence, finding suitable data of refractive index of $VO_2$ for these range of thickness is also a challengeable issue.

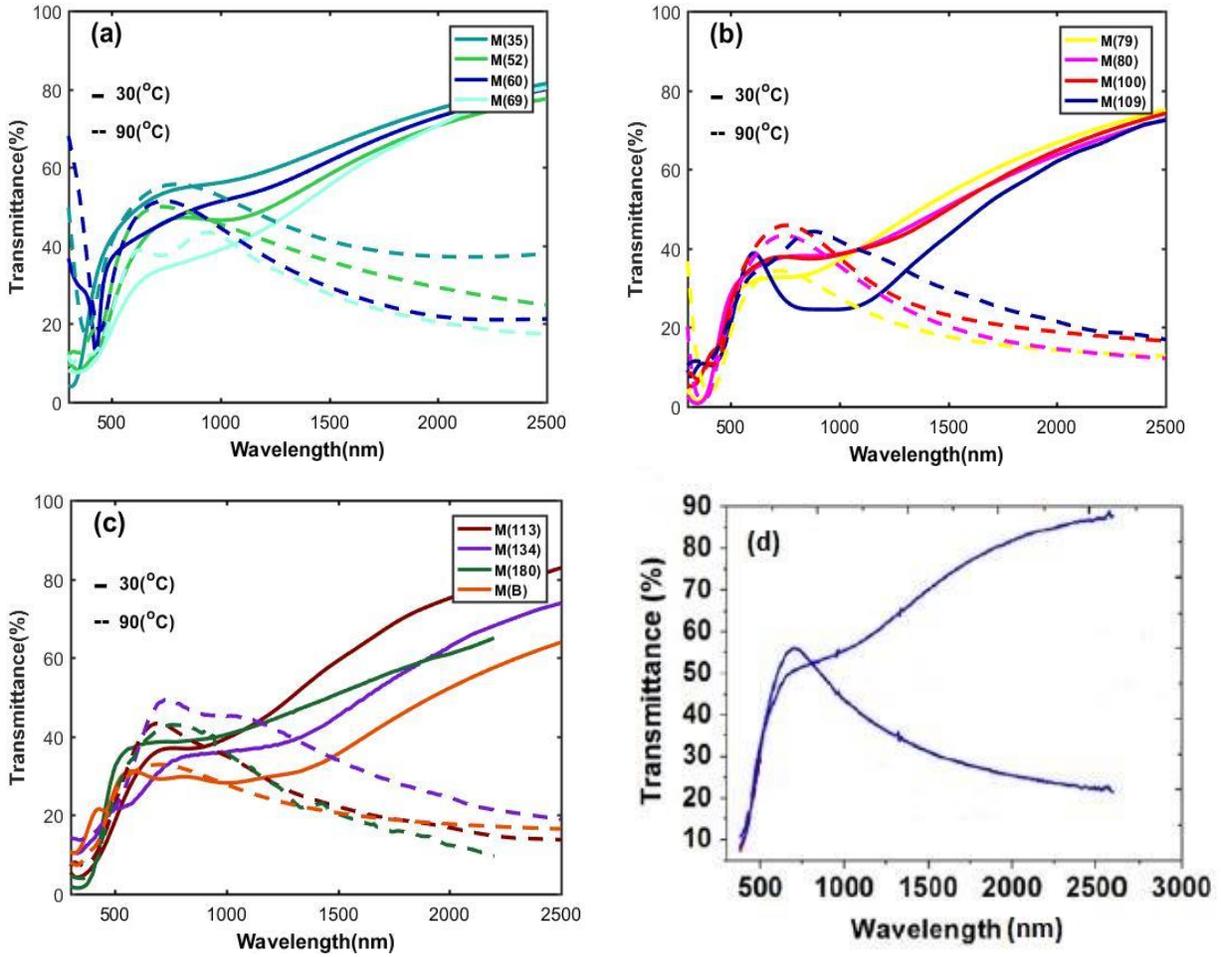

**Fig. 2.** Calculated optical transmittance spectra of 55 nm thick film of $VO_2$ film on glass substrate (fused silica) using refractive index of $VO_2$ films with different thickness **(a)** M(35) to M(69), **(b)** M(79) to M(109), **(c)** M(113) to M(B), and **(d)** experimental data. Semiconductor phase (at $30^0C$) is shown by solid line, and metal phase (at $90^0C$) is shown by dashed line.

The transmission was calculated by using the data in Fig. 1, for refractive index and relation 4. Just as some examples, the calculated transmission for film thicknesses 55 nm, 216 nm, and 540 nm are present in Fig. 2 to 4 (a to c) together with the experimental data (d).

By comparing the calculated data with those of experiments, one can find which data set (labeled by M (x)) can best fit with the experiment. For example, in Fig. 2 to 4, the best fit can be seen for data sets M(60), M(80), and M(35) respectively.

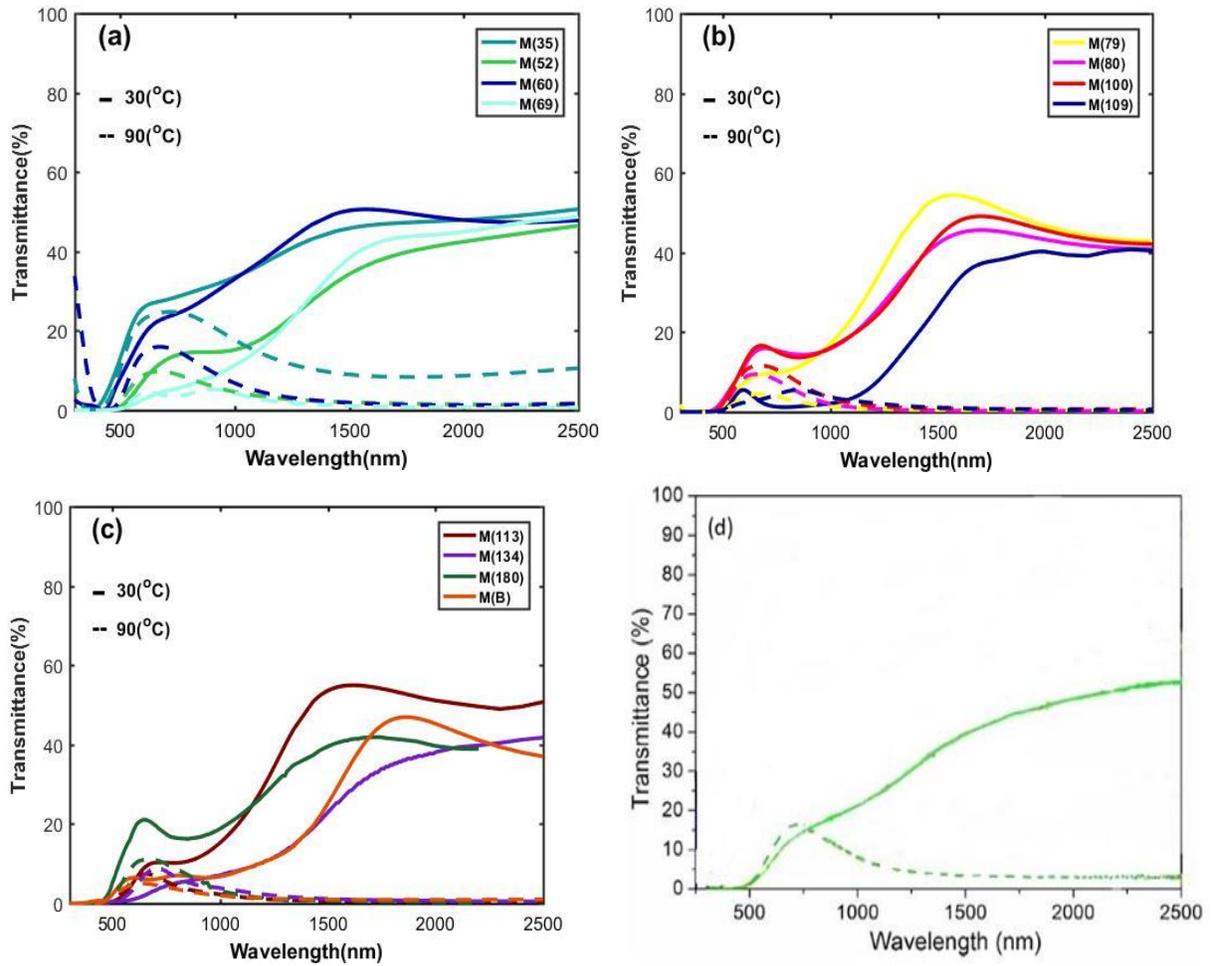

**Fig. 3** Calculated optical transmittance spectra of 216 nm thick film of $VO_2$ film on glass substrate (Soda-lime) using refractive index of $VO_2$ films with different thickness **(a)** M(35) to M(69), **(b)** M(79) to M(109), **(c)** M(113) to M(B), and **(d)** experimental data. Semiconductor phase (at $30^0$C) is shown by solid line, and metal phase (at $90^0$C) is shown by dashed line

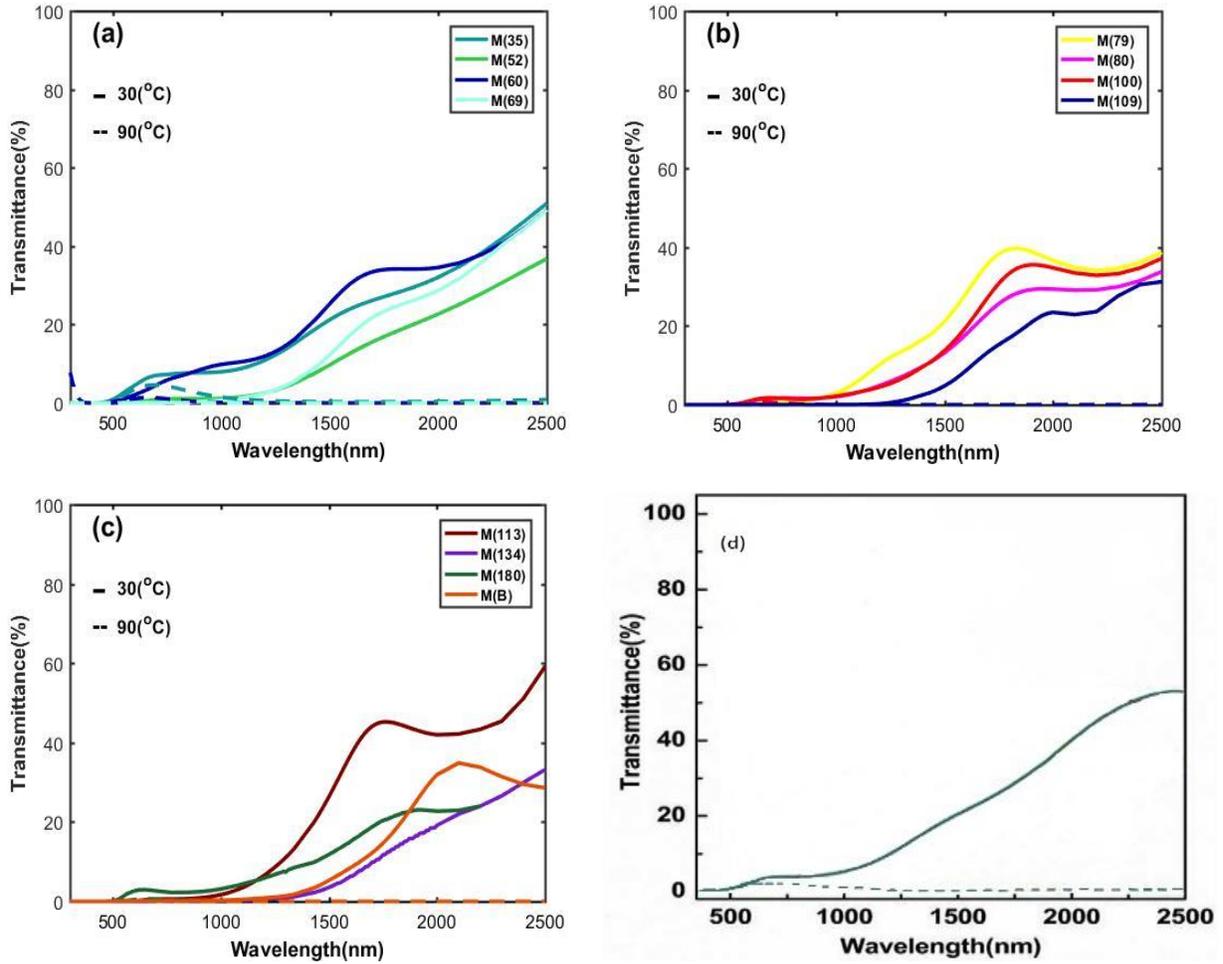

**Fig. 4** Calculated optical transmittance spectra of 540 nm thick film of $VO_2$ film on glass substrate (Borosilicate Crown) using refractive index of $VO_2$ films with different thickness **(a)** M(35) to M(69), **(b)** M(79) to M(109), **(c)** M(113) to M(B), and **(d)** experimental data. Semiconductor phase (at $30^{\circ}C$) is shown by solid line, and metal phase (at $90^{\circ}C$) is shown by dashed line

The parameters $T_{lum}$ and $T_{sol}$ as two important values for evaluating the smart windows performance were also calculated. The data are presented in Fig. 5. This figure shows the above parameters ($T_{lum}$) and ($T_{sol}$) for film thicknesses of 55 nm, 216 nm, and 540 nm. The experimental data are also included in Fig. 5. It can be seen that M(60), M(80), and M(35) are the best data sets which provide the best fit for experimental data and those of the calculations.

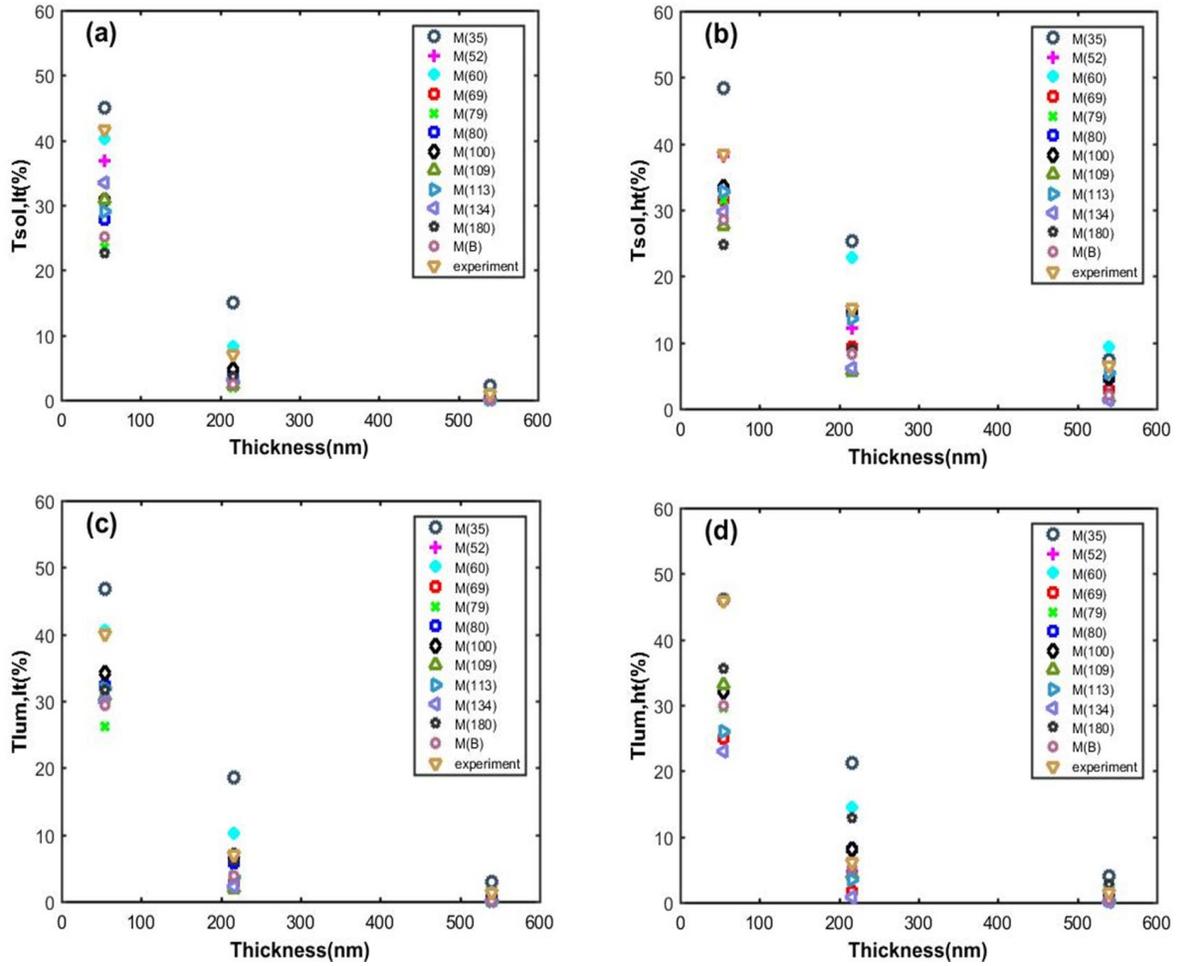

**Fig. 5** Calculated solar and luminous transmittance $T_{lum}$ and $T_{sol}$ for three different thicknesses 55 nm, 216 nm, and 540 nm **(a)** $T_{sol}$ at low temperature ($30^0$C) **(b)** $T_{sol}$ at high temperature ($90^0$C) **(c)** $T_{lum}$ at low temperature ($30^0$C) **(d)** $T_{lum}$ at high temperature ($90^0$C)

The calculation of transmission were performed using all data sets M(x) and the best data sets which confirm the experimental results have been found. The results are summarized in Table 2. The details of these results are presented in Fig. 6.

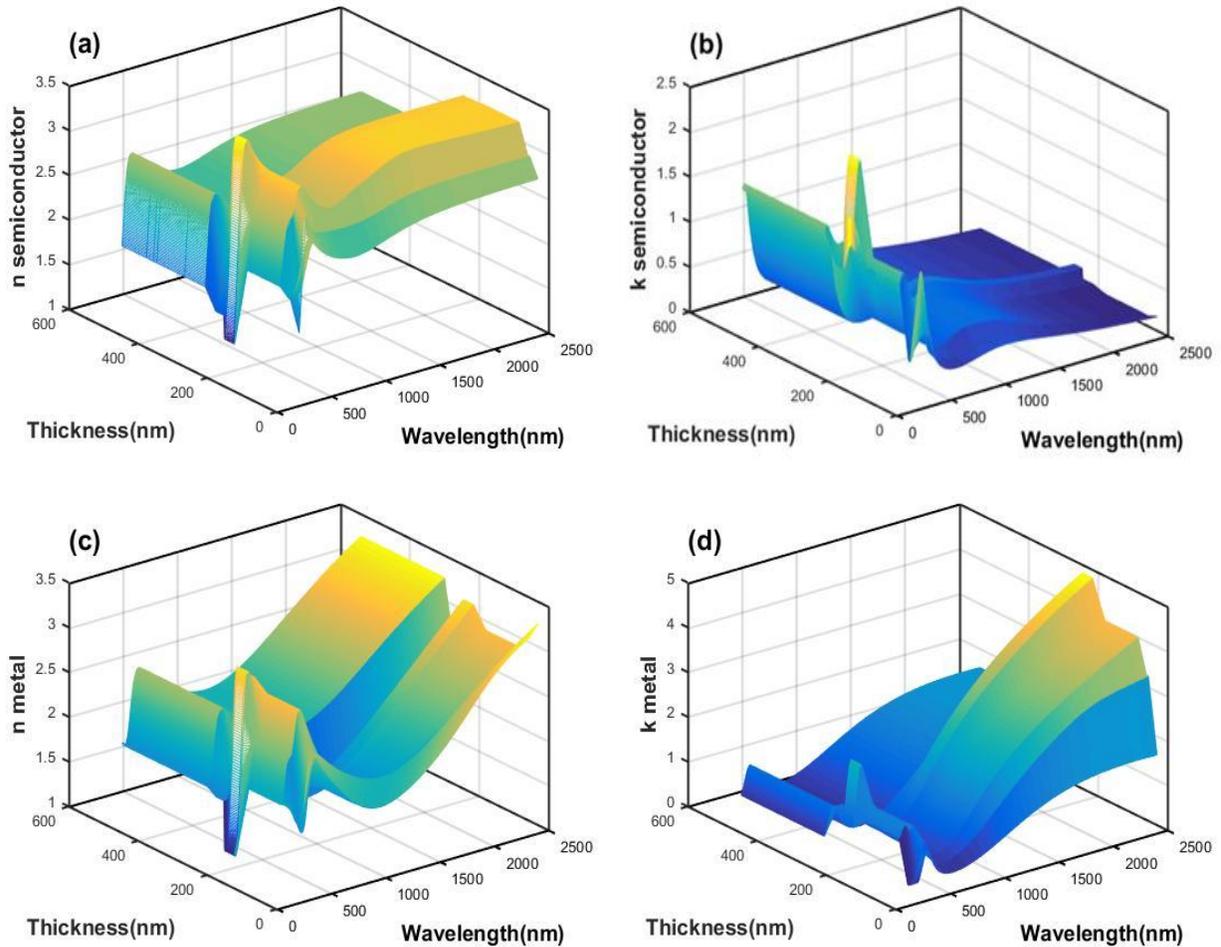

**Fig. 6** Thickness and wavelength dependent refractive index of $VO_2$ film in both semiconductor and metallic states (**a**) real part of refractive index of semiconductor phase (**b**) imaginary part of refractive index at semiconductor phase, (**c**) real part of refractive index of metallic phase, and (**d**) imaginary part of refractive index at metallic phase.

Fig. 6 shows the real and imaginary parts of refractive index of $VO_2$ film in both semiconductor and metallic states which depend on thickness and wavelength. Now, having these results, one can choose the best values for refractive index of $VO_2$ films which can provide any preferred transmission.

**Table 2** The best data sets M(x) that introduced in table 1 which give fairly precise transmission of $VO_2$ films. The $VO_2$ film thickness (nm) are extracted from the references indicated in the middle column.

| $VO_2$ film thickness (nm) | Reference | The best data set M(x) |
|---|---|---|
| 30 | 34 | M(35) |
| 55 | 35 | M(60) |
| 60 | 34 | M(60) |
| 85 | 45 | M(100) |
| 90 | 34 | M(100) |
| 96 | 46 | M(100) |
| 100 | 47 | M(100) |
| 180 | 34 | M(100) |
| 216 | 48 | M(80) |
| 220 | 49 | M(80) |
| 245 | 50 | M(80) |
| 247 | 49 | M(52) |
| 270 | 34 | M(52) |
| 300 | 47 | M(35) |
| 360 | 34 | M(35) |
| 440 | 51 | M(35) |
| 450 | 34 | M(35) |
| 470 | 49 | M(35) |
| 540 | 34 | M(35) |

This question may be raised that how to use these results to calculate transmission in a certain thickness. We classified refractive index of $VO_2$ film "M(x)" based on a specific thickness range and summarized in Table 2. For example, the data set M(100) are more appropriate for calculating the transmission of $VO_2$ film in both semiconductor and metallic within a wide range of thicknesses from 85 nm to 180 nm.

## 4 Conclusion

In this paper we have investigated the problems with vanadium dioxide ($VO_2$) film which is used in thermochromic smart windows. Since, the refractive index of vanadium dioxide film is thickness dependent, in calculating the transmission of light spectrum from $VO_2$ thin films this issue must be considered. To solve this problem, we have used three theoretical models (Lorentz-Drude oscillator, Lorentz oscillator, and Tauc-Lorentz) and employed experimental results from previous experiments to provide thickness dependent refractive index of $VO_2$ films. Using these data sets in calculating transmission of sun light through $VO_2$ films, appropriate thickness dependent refractive index data sets were found. We concluded that the results of this research give new outlook to find a comprehensive relationship for thickness and wavelength dependent refractive index of $VO_2$ film. It was also concluded that the approach in this paper is pretty capable to determine optimized film thickness for the best transmission performance of $VO_2$ films in smart windows. This achievement will help to discover new applications of $VO_2$ not only in smart windows but also in the other optical devices.

*Acknowledgments*

This research did not receive any specific grant from funding agencies in the public, commercial, or not-for-profit sectors.